\documentclass[twocolumn,aps,preprintnumbers,prl,showpacs,%
amsfonts,amsmath,amssymb,superscriptaddress,floatfix]{revtex4}

\usepackage{graphicx,amsmath,amssymb,textcomp}

\newcommand{\affA}{%
	Department of Applied Physics, School of Engineering, 
        The University of Tokyo,\\
	7-3-1 Hongo, Bunkyo-ku, Tokyo 113-8656, Japan}

\newcommand{\affB}{%
	CREST, Japan Science and Technology Agency, 
	1-9-9 Yaesu, Chuoh-ku, Tokyo 103-0028, Japan}
	
\begin{document}

\title{Generation of continuous-wave broadband Einstein-Podolsky-Rosen beams using periodically-poled lithium niobate waveguides}

\date{\today}

\author{Ken-ichiroh Yoshino} 
\author{Takao Aoki}
\author{Akira Furusawa}
\affiliation{\affA}
\affiliation{\affB}

\begin{abstract}

Continuous-wave light beams with broadband Einstein-Podolsky-Rosen correlation (Einstein-Podolsky-Rosen beams) are created with two independent squeezed vacua generated by two periodically-poled lithium niobate waveguides and a half beam splitter. 
The EPR correlation is confirmed with a sufficient criterion 
$\Delta {\rm EPR}= 
\langle \texttt{$[$} \Delta (\hat{x}_{\rm A}-\hat{x}_{\rm B})\texttt{$]$} ^2 \rangle 
+\langle \texttt{$[$} \Delta (\hat{p}_{\rm A}+\hat{p}_{\rm B})\texttt{$]$} ^2 \rangle <1$ and the observed $\Delta {\rm EPR}$ is 0.75 over the bandwidth of 30MHz. 
Although the bandwidth is limited by that of our detector so far, it would be broadened up to 10THz which would be only limited by the bandwidth of phase matching for the second-order nonlinear process.

\end{abstract} 

\pacs{03.67.Hk, 03.67.Mn, 42.50.Dv}

\maketitle

Quantum information processing is intensively studied as next-generation information processing. There are two types of streams for quantum information processing; quantum bit (qubit) \cite{Nielsen00} and continuous variables (CVs)\cite{Sam05}. CV quantum information processing is relatively easy to be implemented in experiments and attracts much interest. The heart of CV quantum information processing is Einstein-Podolsky-Rosen (EPR) correlation \cite{Einstein35} which was originally proposed by Einstein et al. in 1935 \cite{Einstein35}. 

In quantum optics, the EPR correlation is embodied as a two-mode squeezed vacuum \cite{Reid89,Ou92,Wenger05} where quadrature field modes are correlated in two light beams (EPR beams). There are some quantum optical schemes to create them; a continuous-wave (CW) light scheme \cite{Reid89,Ou92} and a pulsed light scheme \cite{Wenger05}. For quantum information processing, the CW scheme might have some advantage, because one can use the well-matured CW scheme of quantum teleportation\cite{Furusawa98,Takei05L} and the spatial and frequency modes of CW EPR beams are well defined and easily used for various quantum information processing. Thus it is very important to create CW EPR beams. 

So far CW EPR beams are generated from optical parametric process with optical-cavity enhancement. The usage of cavities limits the bandwidth of EPR correlation within the cavity bandwidth. For broad band CV quantum information processing, one has to have the other enhancement mechanism of the parametric process. One of the best ways for it is the usage of a waveguide. Since a waveguide can concentrate and keep large power density of a pump beam throughout the interaction length, it does not have any limit of the bandwidth except for that of phase matching.

In this letter, we demonstrate experimental creation of CW EPR beams with two independent squeezed vacua generated by two periodically-poled lithium niobate (PPLN) waveguides and a half beam splitter. 
We observe EPR correlation over the bandwidth of 30MHz. Although the bandwidth is limited by that of the detector so far, it would be broadened up to 10THz which would be only limited by the bandwidth of phase matching for the second-order nonlinear process.

Now let us briefly explain our notation of quantum states including squeezed states and those of EPR beams, and then introduce how to evaluate the quality of EPR beams.
In order to describe the evolution of quantum states of optical field, we use the Heisenberg picture.
We introduce quadrature-phase amplitude operators $\hat{x}$ and $\hat{p}$ corresponding to the real and imaginary part of an optical field mode's annihilation operator $\hat{a} = \hat{x} + i\hat{p}$ (units-free with $\hbar =1/2$, $[\hat{x},\hat{p}]= i/2$).

Squeezed vacua ($\hat{a}_1$, $\hat{a}_2$) in this experiment can be expressed as,
\begin{eqnarray}
\hat{a}_{1} &=& e^{r} \hat{x}_{1}^{(0)} + ie^{-r} \hat{p}_{1}^{(0)},
\nonumber \\ 
\hat{a}_{2} &=& e^{-r} \hat{x}_{2}^{(0)} + ie^{r} \hat{p}_{2}^{(0)},
\end{eqnarray}
where $\hat{x}_{1}^{(0)}$, $\hat{p}_{1}^{(0)}$, $\hat{x}_{2}^{(0)}$, and $\hat{p}_{2}^{(0)}$ are quadrature-phase amplitudes of initial vacuum states and 
$r$ is a squeezing parameter.

By combining these two modes with a half beam splitter (HBS), the two output beams become EPR beams.
Output modes ($\hat{a}_{\rm A}$, $\hat{a}_{\rm B}$) from the HBS can be expressed as,
\begin{eqnarray}
\hat{a}_{\rm A} &=& \hat{a}_{1} + \hat{a}_{2},
\nonumber \\ 
\hat{a}_{\rm B} &=& \hat{a}_{1} - \hat{a}_{2}.
\end{eqnarray}
Thus the output modes ($\hat{a}_{\rm{A}} = \hat{x}_{\rm{A}} + i\hat{p}_{\rm{A}}$, $\hat{a}_{\rm{B}} = \hat{x}_{\rm{B}} + i\hat{p}_{\rm{B}}$) show the following correlation,
\begin{eqnarray}
\hat{x}_{\rm{A}} - \hat{x}_{\rm{B}} &=& \sqrt{2} e^{-r} \hat{x}_2^{(0)}, 
\nonumber \\
\hat{p}_{\rm{A}} + \hat{p}_{\rm{B}} &=& \sqrt{2} e^{-r} \hat{p}_1^{(0)}.
\end{eqnarray}
We check the EPR correlation or entanglement by using the inseparability criterion proposed by Duan \textit{et al}.~\cite{Duan} and Simon~\cite{Simon}.
We define $\Delta \mathrm{EPR}$ described as follows,
\begin{equation}
 \Delta \mathrm{EPR} = \langle [\Delta (\hat{x}_{\mathrm{A}}-\hat{x}_{\mathrm{B}})]^2\rangle +\langle [\Delta (\hat{p}_{\mathrm{A}}+\hat{p}_{\mathrm{B}})]^2\rangle. 
\label{criterion}
\end{equation}
When $\Delta \mathrm{EPR}$ is less than unity, it is proved that these two beams have EPR correlation or are entangled.
In the present experiment we check $\Delta \mathrm{EPR} < 1$  over a wide frequency range.

   \begin{figure}[b]
    \begin{center}
     \includegraphics[width=\linewidth,clip]{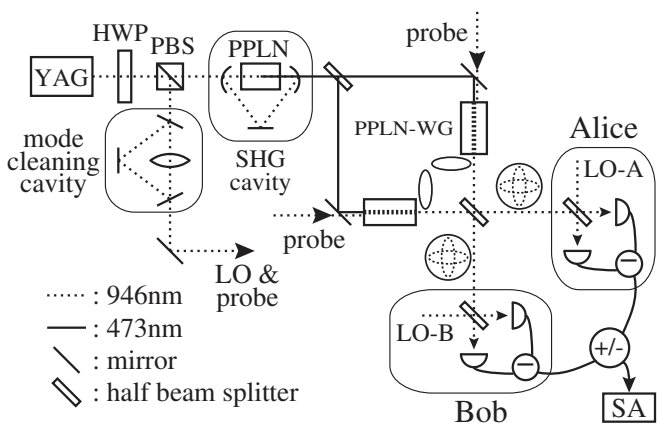}
     \caption{The experimental setup for generation of broadband EPR beams.
HWP: half wave plate, PBS: polarization beam splitter, PPLN-WG: periodically-poled lithium niobate waveguide, LO: local oscillator, SA: spectrum analyzer.
The ellipses indicate the squeezed quadratures of each beam.}
     \label{fig:setup}
    \end{center}
   \end{figure}

Our experimental setup is shown in Fig.\ref{fig:setup}.
Output of a Nd:YAG laser at 946nm (Innolight Mephisto QTL, output power $\sim$ 500mW) is divided into two beams by a half wave plate and a polarization beam splitter.
About 10\% of the total power goes to a mode cleaning cavity (MCC).
The rest 90\% is used for second-harmonic-generation (SHG) to pump PPLN waveguides (NGK corp., Japan) to generate squeezed vacua. 
The shape of cross section of the PPLN waveguides is a trapezoid whose base and height are about 5 and 3.5 $\mu$m, respectively. The length of the waveguide is 12 mm. The squeezed vacua and the EPR beams are measured with homodyne detectors which have Si photodiodes (Hamamatsu S3590-06, anti-reflective coated at 946nm) with quantum efficiency of 99.4\% at 946nm.

The MCC prepares local oscillators (LOs) for homodyne detection.
The MCC acts as a spatial filter, so we can obtain a nearly ideal Gaussian fundamental mode.
Moreover, we can obtain very stable output power from the MCC because the cavity length is so controlled that the output power stays to be constant, instead of locking to a peak of cavity transmission.
As the result, we can reduce the drift of the LO powers within 0.2dB.

About 90\% of the output of the Nd:YAG laser is frequency-doubled in an external  cavity with a bulk crystal of PPLN inside.
At the early stage of the experiment, we tried to use a PPLN waveguide for second harmonic generation (SHG), but the coupling efficiency of fundamental wave (946nm) to the waveguide was very low (about 65\% at most), and actual SHG efficiency of the waveguide was worse than that of the external cavity with a bulk PPLN crystal inside.
Therefore we chose a bulk PPLN crystal for SHG, and it can generate 190mW second harmonic at 473nm from 430mW of fundamental power.
The SHG output is divided into two beams by a half beam splitter to pump the PPLN waveguides.
About 90mW second harmonic at 473nm goes to each PPLN waveguide, and about 30mW can be coupled.
Broadband squeezed vacuum states are generated by parametric amplification process in the waveguides, which is the reverse process of SHG.

By using phase matching condition of a PPLN waveguide, we can calculate the phase matching linewidth~\cite{spectrum1,spectrum2}.
According to the calculation, the linewidth of the squeezed vacua generated by the PPLN waveguides is about 30nm, and it corresponds to about 10THz bandwidth in the frequency domain.
This means the observed bandwidth of squeezed vacua from the PPLN waveguides is only limited by that of the homodyne detectors ($\sim$ 30MHz).

The EPR beams are created by combining two squeezed vacua from each PPLN waveguide at a half beam splitter with orthogonal phases ($\pi /2$ phase shift) to each other as shown by the ellipses in Fig.\ref{fig:setup}.
For this purpose, we inject weak coherent beams from the MCC to the waveguides, which are denoted as `probes' in Fig.\ref{fig:setup}. The phases of the coherent beams are locked to the squeezing quadratures. With the interference fringe between the coherent beams at the half beam splitter, we lock the relative phase of the squeezed vacua. Furthermore, the coherent beams have modulation sidebands at 70 kHz and  80 kHz, respectively and they are used for locking the LO phases at the homodyne detectors.

The output beams of the half beam splitter are distributed to Alice and Bob's homodyne detectors, respectively, and they measure noise powers of the beams in each quadrature-phase by a spectrum analyzer (Agilent E4401).
Spatial mode matching efficiencies between the coherent beams and LOs, which represent those between squeezed vacua from the waveguides and LOs, are about 0.94 and 0.86, respectively.
Alice and Bob's LO powers are set to be 3.5mW, and they are stabilized by the MCC as mentioned before.

   \begin{figure}[tb]
    \begin{center}
     \includegraphics[width=\linewidth,clip]{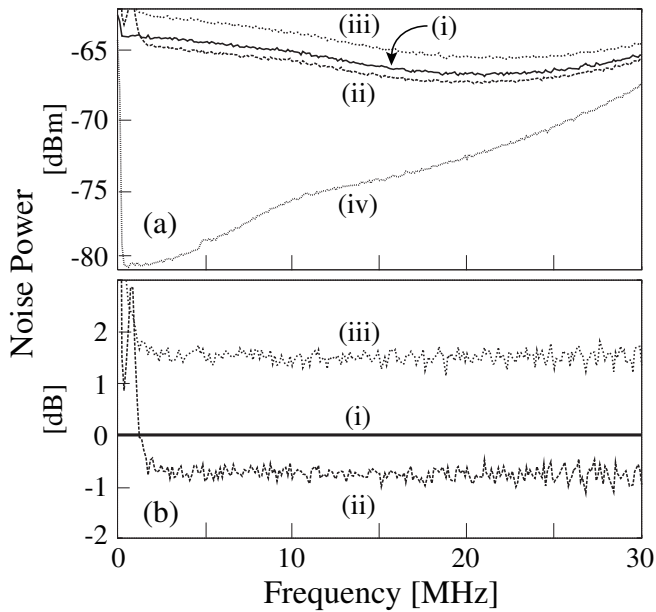}
     \caption{Noise measurement results on a squeezed vacuum generated by a PPLN waveguide.
These are recorded by Alice's homodyne detector after passing through a half beam splitter.
(a) is the raw results including the detector's noise. (b) is the results after compensation. The vacuum noise level is normalized to 0dB and the effect of the detector's noise is subtracted.
In each figure, traces (i), (ii), (iii) show the noise levels of a vacuum, squeezed quadrature, and anti-squeezed quadrature respectively.
In figure (a), trace (iv) represents the detector's noise.
The LO power is set to 3.5mW, resolution bandwidth is 100kHz, and video bandwidth is 100Hz.
All traces are averaged ten times.}
     \label{fig:SQZ}
    \end{center}
   \end{figure}

First, we measure a squeezed vacuum from one PPLN waveguide at the Alice's homodyne detector.
Here, the other squeezed vacuum is blocked just after the other PPLN waveguide.
Fig.\ref{fig:SQZ} shows the results measured by a spectrum analyzer.
Since the squeezed vacuum is measured after passing through the half beam splitter, it suffers from 50\% losses.

Fig.\ref{fig:SQZ}(a) shows raw data on noise powers of a vacuum, the squeezed quadrature, the anti-squeezed quadrature and the homodyne detector itself.
In Fig.\ref{fig:SQZ}(b), the noise level of the vacuum state is normalized to 0dB, and the detector noises are subtracted.
The noise level of the squeezed vacuum is about $-$0.76dB compared to that of the vacuum, and it is almost flat up to 30MHz.
If the squeezed vacuum would be measured directly, that is, without the half beam splitter, the squeezing level would be calculated as $-$1.7dB.

Next, we proceed to the measurement of EPR correlation 
and unblock the other squeezed vacuum.
We set the LO phases at both of the homodyne detectors so as to obtain $x$ or $p$ quadrature component of each beam.
The signals from the homodyne detectors are subtracted or summed electronically and the variance is measured by a spectrum analyzer in a wide spectral range.
In order to get the value of
$\langle [\Delta (\hat{x}_{\mathrm{A}}-\hat{x}_{\mathrm{B}})]^2\rangle$,
we set the LO phases at homodyne detectors to $x$ quadrature, and for 
$\langle [\Delta (\hat{p}_{\mathrm{A}}+\hat{p}_{\mathrm{B}})]^2\rangle$ 
we set it to $p$ quadrature.
Then $\Delta \mathrm{EPR}$ is calculated with eq.(\ref{criterion}).

   \begin{figure}[tb]
    \begin{center}
     \includegraphics[width=\linewidth,clip]{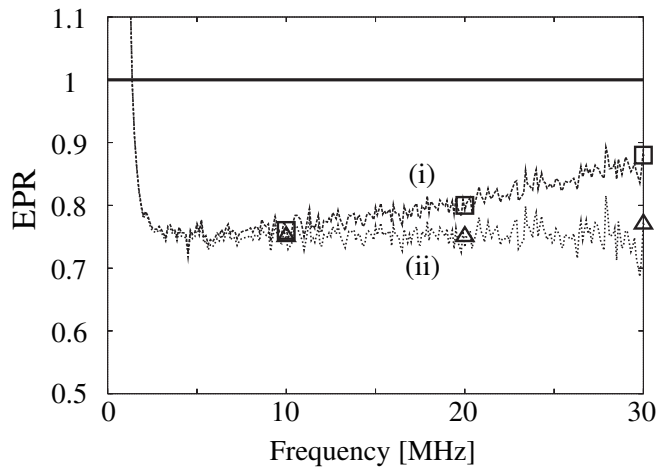}
     \caption{Results on the inseparability criterion,
$\Delta \mathrm{EPR}=\langle [\Delta (\hat{x}_{\mathrm{A}}-\hat{x}_{\mathrm{B}})]^2\rangle +\langle [\Delta (\hat{p}_{\mathrm{A}}+\hat{p}_{\mathrm{B}})]^2\rangle$.
(i) and (ii) represent $\Delta \mathrm{EPR}$ at 100kHz RBW before and after subtracting the effects of detector's noise respectively.
Squares and triangles are results at 5MHz RBW before and after noise compensation.}
     \label{fig:Duan}
    \end{center}
   \end{figure}

The experimental results on $\Delta \mathrm{EPR}$ are shown in Fig.\ref{fig:Duan}.
We can see that $\Delta \mathrm{EPR}$ is below unity up to 30MHz, which is the detector bandwidth.
Furthermore, if the detector noises are subtracted, $\Delta \mathrm{EPR}$ is nearly flat at about 0.75 over the whole spectral range (curve (ii)).

Up to here, we took a narrowband spectral filter of the spectrum analyzer (resolution bandwidth RBW = 100kHz) in order to see the frequency characteristics in detail.
However, since our present goal is to broaden the bandwidths of EPR correlation, we measure $\Delta \mathrm{EPR}$ with a broadband filter (RBW=5MHz), which is the broadest filter of the spectrum analyzer.
The results are plotted in Fig.\ref{fig:Duan}.
Even when the RBW is broadened, we can still see almost the same value of $\Delta \mathrm{EPR}$.
It indicates that broadband EPR beams are generated successfully. 
Note that although we use the weak coherent beams to lock the phases of squeezed vacua and LOs in the present experiments, we can remove it by using the `sample and hold' technique \cite{Takei06}.

In conclusion,
CW EPR beams are created with two independent squeezed vacua generated by two PPLN waveguides and a half beam splitter. 
The EPR correlation is confirmed with a sufficient criterion 
$\Delta {\rm EPR}= 
\langle \texttt{$[$} \Delta (\hat{x}_{\rm A}-\hat{x}_{\rm B})\texttt{$]$} ^2 \rangle 
+\langle \texttt{$[$} \Delta (\hat{p}_{\rm A}+\hat{p}_{\rm B})\texttt{$]$} ^2 \rangle <1$ and the observed $\Delta {\rm EPR}$ is 0.75 over the bandwidth of 30MHz. 
Although the bandwidth is limited by that of our detector so far, it would be broadened up to 10THz which would be only limited by the bandwidth of phase matching for the second-order nonlinear process.

\end{document}